# Strong Circular Dichroism in Single Gyroid Optical Metamaterials

Cédric Kilchoer[1], Narjes Abdollahi[1], James A. Dolan[1], Doha Abdelrahman[1], Matthias Saba[1], Ulrich Wiesner[2], Ullrich Steiner[1], Ilja Gunkel[1] and Bodo D. Wilts[1*]

Over the past two decades, metamaterials have led to an increasing number of biosensing and nanophotonic applications due to the possibility of a careful control of light propagating through subwavelength features. Chiral nanostructures (characterized by the absence of any mirror symmetry), in particular, give rise to unique chiro-optical properties such as circular dichroism and optical activity. Here, we present a gyroid optical metamaterial with a periodicity of 65 nm exhibiting a strong circular dichroism at visible wavelengths. Our bottom-up approach, based on metallic replication of the gyroid morphology in triblock terpolymer films, generates a large area of periodic optical metamaterials. We observe a strong circular dichroism in gold and silver gyroid metamaterials at visible wavelengths. We show that the circular dichroism is inherently linked to the handedness of the gyroid nanostructure, and demonstrate its tuneability. The optical effects are discussed and compared to other existing systems, showing the potential of bottom-up approaches for large-scale circular filters and chiral sensing.

## 1 Introduction

Three-dimensional (3D) chirality, i.e. the absence of mirror or inversion symmetry, can be found in a large number of natural morphologies including proteins and various crystal structures [1]. Light propagation in a chiral medium depends on its polarization state, leading to chiro-optical effects such as circular dichroism (CD) and optical activity [1]. Traditionally, CD is the difference in absorption between circularly polarized light of opposite handedness, while optical activity refers to the rotation of linearly polarized light propagating through a chiral medium due to a difference in phase velocity, also known as circular birefringence [2]. These two effects are typically weak in natural materials, but they can be strongly enhanced in specifically engineered nanostructured dielectric [3] and plasmonic composites [4]. The latter are better known as chiral metamaterials and possess a subwavelength structure, which furnishes them with optical properties that are often not found in nature. Optical metamaterials – nanostructured materials engineered to yield a desired response in the visible spectrum – have received special attention due to exciting potential nanophotonic applications including plasmonic biosensing [5], and subdiffraction imaging [6]. In addition, chiral nanostructures can strongly enhance both optical activity and CD [4], leading to attractive effects at visible frequencies with chiral metamaterials serving as highly efficient broadband circular polarizers [7] and chiral sensors [8, 9]. Ultimately, a strong chirality may gives access to negative refractive indices as predicted by Tretyakov [10] and Pendry [11]. 3D negative refraction may lead to a number of technological applications, most notably the 'perfect lens' [12], and the invisibility cloak [13]. Chiral metamaterials with an optical response at visible and near-infrared frequencies are particularly challenging to fabricate because of the required 3D chiral structures on the nanoscale [14], in contrast to their counterparts at longer wavelengths [15–17]. Planar chiral periodic nanostructures such as gammadions, fabricated by conventional lithography techniques, exhibit a CD at their respective resonance frequencies [18–20]. The strength of the CD is however limited by the two-dimensional nature of the structure. By stacking and twisting several layers of planar metamaterials, the CD can be enhanced and extended to a broader frequency range, both at infrared [21–24] and optical wavelengths [25]. However, the sequential lithographic manufacture of stacked morphologies is time-consuming and the alignment of successive layers can be demanding. Instead of stacking planar metamaterials, CD can alternatively be generated in metamaterials that feature three-dimensional chiral elements [2]. Generally, 3D chiral metamaterials can be fabricated either top-down, or by using self-assembly bottom-up techniques. Top-down fabrication mostly focused on helical structures arranged in 2D arrays, which often show a strong CD [7, 26]. For example, Gansel *et al.* demonstrated periodic gold helices exhibiting a strong CD at mid-infrared wavelengths in transmission by employing direct laser writing followed by gold electrodeposition [7]. These helical structures and their CD can be further optimized by increasing the number of intertwined helices within the unit cell [27–29]. Employing focused ion beam-induced deposition techniques, helices with a smaller radius were fabricated, showing a strong CD between 500 and 1000 nm [30]. By carefully controlling the deposition angle, helices can be grown from gold nanoparticle seeds yielding sub-100 nm gold helices with a material-based tuneable CD at visible frequencies [31, 32]. Existing top-down techniques are often cumbersome, costly, and limited in the achievable feature sizes and coverage areas. On the other hand, bottom-up manufacture of chiral metamaterials has virtually no limitations in the feature size or morphology. While metamaterials based on self-assembly techniques such as DNA origami [33–35], peptides [36, 37] and cellulose nanocrystals templates [38] show CD at visible wavelengths, its strength is several orders of magnitudes lower compared to helical metamaterials [4, 9].

[1] C. Kilchoer, N. Abdollahi, Dr. J. A. Dolan, D. Abdelrahman, Dr. M. Saba, Prof. U. Steiner, Dr. I. Gunkel, Dr. B. D. Wilts
Adolphe Merkle Institute, University of Fribourg, Ch. des Verdiers 4, 1700 Fribourg, Switzerland
*E-mail: bodo.wilts@unifr.ch
[2] Prof. U. Wiesner
Cornell University, 214 Bard Hall, Ithaca, NY 14853-1501, USA

Here, we present an optical metamaterial fabricated by replication of a self-assembled gyroid in block copolymer (BCP) films [39, 40] exhibiting strong CD at visible frequencies. Triblock terpolymer films with an alternating gyroid morphology exhibit an intricate nanostructure with a cubic unit cell of approximately 65 nm [41]. Large single gyroid gold domains, inclined along their cubic $\langle 110 \rangle$ direction, display a linear polarization-dependent optical response [39, 41, 42], which has only very recently been explained [41]. While the same sample displayed no discernible CD when illuminated at normal incidence, a weak *gyrotropic* effect was observed when the sample was tilted to align its $\langle 111 \rangle$ axis with the illumination direction [39].

In this article, using the same gyroid morphology, we demonstrate a strong CD in silver and gold gyroid metamaterials with a peak at ∼510 nm and ∼550 nm, respectively. We further show that the sign of the CD is linked to the handedness of the gyroid structure and can be tuned using different metals (Au, Ag) and dielectric surrounding materials. The CD strength in these self-assembled gyroid metamaterials is comparable to the best results achieved with top-down fabrication techniques [4], demonstrating the viability of self-assembly in fabricating metamaterials with strong chiro-optical responses. We thus show that our bottom-up approach is a promising, fast, and cheap alternative to established top-down techniques.

## 2 Results

Continuous nanostructured gold and silver networks exhibiting the alternating gyroid morphology (body-centered cubic space group $I4_132$, 214 in [43]) were fabricated by a procedure that is discussed in detail in a separate publication [44]. In brief, the fabrication starts with the self-assembly of a polyisoprene-*b*-polystyrene-*b*-poly(ethylene oxide) BCP film into the alternating gyroid morphology. The polyisoprene (PI) gyroid phase is degraded, followed by electrochemical backfilling of the voided single gyroid network with gold or silver (**Figure 1**a) [40]. The (in–plane) single gyroid morphology of the polymer templates and the metal replica was previously confirmed by small–angle X–ray scatter-ing [40]. We note that grazing-incidence small-angle X-ray scattering of similar gyroid terpolymer films revealed a slight distortion in the out–of–plane direction, which is related to the film processing [45]. Importantly, the subsequent processing of the gold and silver gyroid manufacture differ. In the case of gold gyroids, the remaining polymer scaffold is etched away in an oxygen plasma, yielding a free-standing single gyroid gold film [40]. Because of the lower oxidation resistance of silver compared to gold, this approach does not yield metallic silver gyroids. Instead, polymer degradation is carried out in an argon plasma for the manufacture of free-standing silver gyroids. As explained in detail in [44], this results in stable metallic single silver gyroids that are covered by a very thin carbonaceous coating.

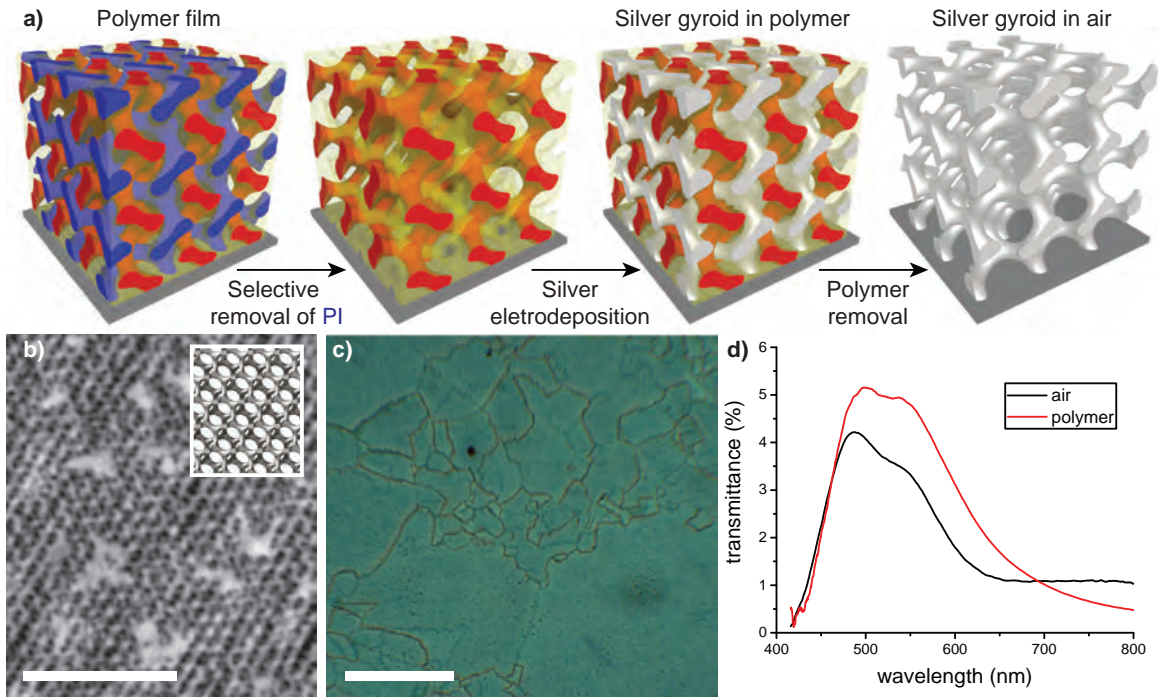

**Figure 1.** Fabrication and optics of silver single gyroids. a) Sketch of the fabrication process of single gyroid optical metamaterials. b) SEM top-view image of a silver gyroid metamaterial. The gyroid structure exhibits a ⟨110⟩ out-of-plane orientation. Inset: Schematic gyroid structure viewed from the ⟨110⟩ direction. c) Multi-domain gyroid optical metamaterial in transmission under unpolarized light at normal incidence. Large domains appear green, while the borders between the domains are clearly visible by their darker coloration. d) Unpolarized transmission spectrum of a single silver gyroid domain with and without polymer. Scale bars: (b) 500 nm, (c) 200 µm.

We first focus on silver gyroid metamaterials that have a stronger optical activity compared to gold. Top-view SEM images show that the fabricated free-standing silver single gyroid has a periodic unit cell of circa 65 nm and a ⟨110⟩ out-of-plane orientation (Figure 1b), in agreement with previous work using the same triblock terpolymer [40, 41]. The periodicity the gyroid morphology was confirmed throughout the entire film thickness by focused ion beam SEM (Figure S3). Transmission optical micrographs of silver gyroid films on FTO-coated glass substrates reveal large green-coloured domains (Figure 1c). These correspond to single domains with uniform gyroid orientation (Figure 1b, see also ref. [40]). The green color stems from a transmission band in the 450 to 600 nm wavelength range, with a peak transmittance of ∼0.045 at 500 nm (Figure 1d).

### 2.1 Circular dichroism of single gyroid domains

SEM images of the nanostructure (**Figure 2**a) show the presence of single gyroid domains of different handedness. Since our fabrication method does not induce a unique handedness, a racemic coexistence of both handednesses across the sample is expected. Upon illumination with circularly-polarized (CP) light, multidomain silver gyroid films show a mosaic of bimodally colored domains (Figure 2b,c). The color and intensity of individual domains swaps when the handedness of the light is altered. The difference in transmittance of individual domains when illuminated under right-handed (RCP) and left-handed (LCP) circularly polarized light (defined from the receiver's point of view) indicates the presence of a strong CD, which was spectrally quantified using a custom-built microspectrophotometer with a measurement spot size diameter of ∼50 µm. This measurement allows the characterization of individual domains that typically span a few hundred µm. Figure 2d shows a pronounced difference in the transmitted intensity between 450 and 550 nm. As common for photonic crystals [47] and metamaterials [23], we define the CD as

$$\text{CD} = \frac{T_{\text{RCP}} - T_{\text{LCP}}}{T_{\text{RCP}} + T_{\text{LCP}}}, \tag{1}$$

where $T_{\text{RCP}}$ and $T_{\text{LCP}}$ denote the transmitted intensity for RCP and LCP illumination, respectively. Note that the CD in this definition is generally a combination of two effects: a difference in absorption of RCP and LCP light as in the traditional definition [1], and a different distribution between reflection and transmission for the two polarizations as in photonic crystals [3].

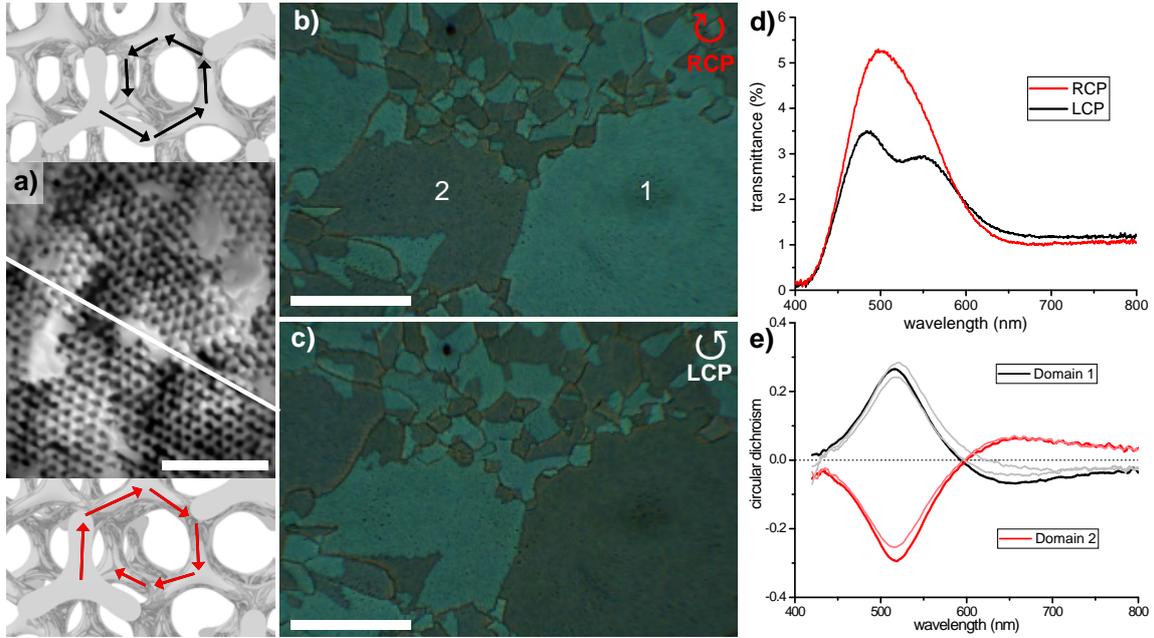

**Figure 2.** Circular dichroism in a silver gyroid optical metamaterial. (a) SEM image of a single gyroid nanostructure, encompassing a grain boundary, indicated by the white line. To identify the handedness of single gyroid, the [110]-oriented sample was tilted by 35° around the [1$\bar{1}$0] axis, providing a view of the [111] orientation [46]. The right- and left-handed domains at the top and bottom of the image are illustrated above and below the SEM image, respectively. Illumination of the [110]-oriented silver gyroid film with (b) right-handed and (c) left-handed circularly polarized light shows a mosaic of bimodally colored domains, where the contrast is inverted with handedness. (d) The CP-dependent transmittance spectra of an individual domain (domain 1) shows a strong difference. (e) CD measurements of 7 different domains show either positive or negative spectra of identical shape. Domains 1 and 2 are indicated in (b). Scale bars: (a) 500 nm, (b,c) 200 µm.

The CD of different domains is plotted in Figure 2e. While the overall strength of CD is uniform across different domains (maximal absolute value of ∼0.26 at ∼510 nm for all measured domains), it is either positive or negative indicating a bimodal behavior. This implies that the CD arises from the inherent chirality of right- and left-handed single gyroid network domains.

## 2.2 Influence of the plasmonic metal and dielectric medium

To determine the effect of the plasmonic metal, from which the gyroid is formed, on the CD, we compare the optical responses of single gyroid networks made from gold and silver. **Figure 3**a shows the CP-dependent transmittance of gold and silver single gyroid networks. The gold spectra (solid lines) show two transmittance peaks at ∼490 nm and 580 nm and a dip at ∼520 nm, as previously observed [48, 49]. While the short-wavelength peak is nearly insensitive to the CP handedness, the intensity and wavelength of the second peak and the dip differ. The maximum CD of $\sim 0.04$ is seen at $\sim 550\,\text{nm}$ (Figure 3b). In silver gyroid metamaterials, the transmittance arising from the resonant nanostructure is spectrally more confined, at wavelengths between 450 and 600 nm. A dip at $\sim 510\,\text{nm}$ is present for one of the two handednesses. $T_{\text{RCP}}$ and $T_{\text{LCP}}$ show a pronounced difference that leads to a much stronger CD with a peak value above 0.25 at 510 nm, 5-6 times higher compared to the gold metamaterial.

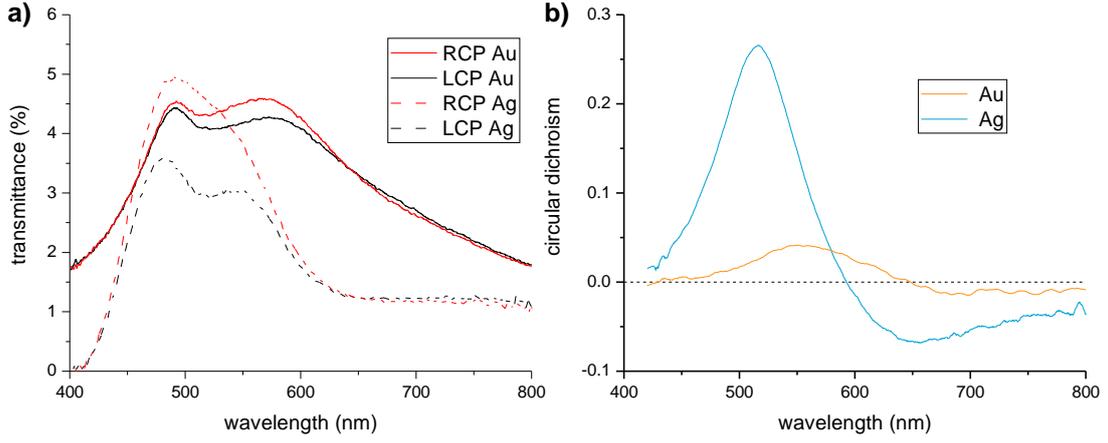

**Figure 3.** Circular polarization resolved transmittance. (a) CP-dependent spectra of gold and silver single gyroid metamaterials. (b) The CD of gold sample is red-shifted and has a lower intensity compared to the silver metamaterial.

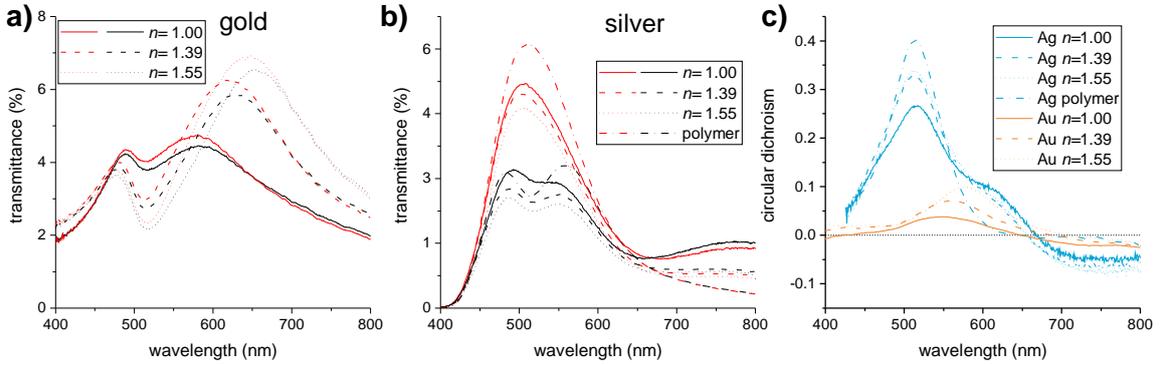

**Figure 4.** Effect of the dielectric medium on the CD. Measured CD for (a) gold and (b) silver single gyroid metamaterials before removing the polymer scaffold ($n \approx 1.6$) and after infiltration of the gyroids with several dielectric media (air $n \approx 1.00$, heptane $n \approx 1.39$, Cargille refractive index liquid $n \approx 1.55$). (c) Wavelength-dependent CD intensity as a function of the refractive index of the dielectric medium.

The optical response of nano-structured plasmonic metals is known to be sensitive to the environment, i.e. the refractive index of the dielectric medium surrounding the plasmonic features [48]. To study this effect, the silver gyroids were measured before the polymer matrix was etched away ($n \approx 1.60$) and, furthermore, the voided gold and silver gyroid metamaterials were infiltrated with several dielectric media. **Figure 4**a,b compares the CD of gold and silver single gyroids infiltrated by air ($n = 1.00$), heptane ($n = 1.39$), and Cargille refractive index liquid ($n = 1.55$). In gold (Figure 4a), the second transmission peak is enhanced and red-shifted, and the dip is increasingly pronounced with increasing dielectric refractive index. This leads to a red-shift and a strong increase in the CD peak value, shifting from $\sim 0.04$ at 550 nm in air to $\sim 0.10$ at 575 nm for a dielectric refractive index of 1.55 (Figure 4c orange curves).

The transmission spectra for the silver gyroid differ significantly from those made from gold. A cursory investigation of the spectra in Figure 4b reveals that they are spectrally invariant with respect to the dielectric environment. The strong short-wavelength decay of the spectra is noteworthy, since none of the materials in the sample exhibit absorption in this wavelength region. Secondly while the gold spectra exhibit a double-peak structure in both polarization channels, a similar spectral shape is seen only in the LCP silver spectra.

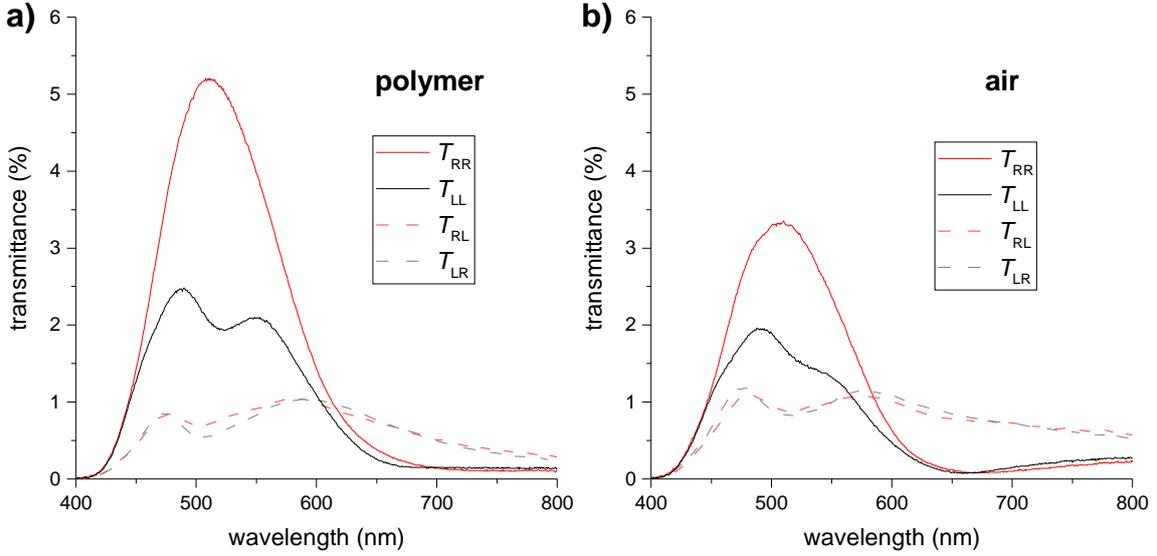

**Figure 5.** Polarization conversion efficiency of silver gyroids in air and embedded in the polymer scaffold. Transmittance spectra of the four elements of the Jones matrix, where $T_{\text{RL}}$ and $T_{\text{LR}}$ are measures for the conversion of CP light by silver single gyroid metamaterials surrounded by (a) polymer and (b) air. Gold gyroid metamaterials were exhibiting a extremely weak polarization conversion, therefore this result in not shown.

A more careful analysis of the spectra in Figure 4b reveal two opposing trends in the presence of dielectric media. In the presence of the polymer scaffold, the silver is surrounded by the polystyrene block of the BCP ($n \approx 1.60$), which, similar to the gold spectra, enhances transmission, particularly in the RCP channel. The voided samples show a much reduced transmission, which weakly decreases in both channels with increasing refractive index of the infiltrated dielectric liquids in silver gyroids. The strong difference of the RCP transmission at 500 nm between the polymer-containing sample and the sample infiltrated with the $n = 1.55$ liquid suggests that the etching process, which removes the polymer, significantly changes the immediate environment of the silver gyroid struts.

Similar to the Au data, the CD is strongly enhanced for the polymer-containing silver gyroid sample. While the liquid-infiltrated silver gyroid samples also exhibit a CD increase compared to the voided air gyroid, this has to be seen in context of an overall transmission reduction in both CP channels for these two samples.

### 2.3 Polarization conversion in silver gyroids

To gain an understanding how different dielectric background media influence the CD strength and spectral position, the optical response was analyzed by two sets of a linear polarizer and a quarter wave plate, one each in the illumination and detection optical paths. This optical configuration allows the quantification of the conversion of the circular polarization by the silver gyroid samples. **Figure 5** shows the presence of polarization conversion between LCP and RCP light. The conversions from right-to-left ($T_{\text{RL}}$) and from left-to-right ($T_{\text{LR}}$) circularly polarized light have similar intensities, so that the CD evidently arises from a difference between $T_{\text{RR}}$ and $T_{\text{LL}}$. This behaviour is expected as the geometry exhibits a $C_2$ rotational symmetry perpendicular to the optical axis for light at normal incidence on a $\langle 110 \rangle$ inclined gyroid (cf. *generalized chiral structures* in [50]). We note that the above statement additionally requires reciprocal materials and that the slab geometry is terminated such that the bulk $C_2$ axis lies in its center. The former requirement is clearly satisfied, while the latter is only approximately valid in our self-assembled optical metamaterials.

For the free-standing silver gyroid (Figure 5a), the majority of the circularly polarized light is transmitted without polarization conversion ($T_{\text{RR}}$ and $T_{\text{LL}}$) below 570 nm. Part of the light is, however, converted into light of the other handedness (compare dashed lines with the solid lines). For a gyroid with the polymer scaffold in place ($n = 1.60$, Figure 5b), the light transmitted without polarization conversion is increased by approximately 30%, while the polarization conversion is comparable to that of the voided gyroid. In both materials, the polarization conversion has a minimum around 510 nm, where the CD is strongest. Over the whole spectral range, silver gyroid embedded in polymer shows a much lower ratio of polarization conversion compared to air. This difference in the ratio of polarization conversion explains the weaker CD in silver gyroid surrounded by air. Interestingly, in both cases, the converted light becomes dominant at longer wavelengths, where up to 90% (at at 670 nm in air) of the transmitted light is converted into the other handedness.

Chiral metamaterials can be extremely efficient in converting the polarization state of incident light [51–53]. Due to the lack of a higher rotational symmetry ($C_n$ with $n > 2$, [3]) in $\langle 110 \rangle$ oriented gyroids, light propagation in the metamaterial is not restricted to one polarization channel. As a result, RCP light can be partially converted into LCP light and vice-versa. The effect of bulk and slab symmetries on the chiro-optical response will be discussed below.

## 3 Discussion

### 3.1 Circular dichroism in single gyroid nanostructures

Silver and gold single gyroid metamaterials show a strong and variable CD in the visible wavelength range. Gyroids are intrinsically chiral, although their chiro-optical response strongly depends on the observation direction if realized as a photonic crystal [54] or a microwave metamaterial [55]. The gyroid handedness is defined by the rotational sense of the 6-fold helix along the $\langle 111 \rangle$ axis (Figure 2a). The origin of the CD in plasmonic single gyroids has been theoretically studied by Oh *et al.* for gyroids which are oriented along the $\langle 100 \rangle$ and $\langle 111 \rangle$ directions [55, 56]. For a perfect electric conductor, a right-handed single gyroid network has a negative CD when oriented along the $\langle 100 \rangle$ direction. For the $\langle 111 \rangle$ direction, the CD is virtually absent [55]. Theory suggests that the CD vanishes in any loss-less metamaterial with an inherent $C_3$ axis along the optical axis, *i.e.* the $\langle 111 \rangle$ direction of a gyroid, in a converged simulation [3]. The fabricated single gyroid metamaterial investigated here is however $\langle 110 \rangle$ out-of-plane oriented, for which a theoretical description is yet to be performed.

The single gyroid optical response is experimentally characterized by a strong CD (Figure 2), with a finite circular polarization conversion (Figure 5). These findings can be associated with the metamaterial symmetry properties. In terms of symmetry, the gyroid morphology belongs to the non-symmorphic space group $I4_132$ [57], with underlying isogonal octahedral $O$ point group. For the gyroid slab geometry, oriented along the cubic $\langle 110 \rangle$ direction, this is reduced to the dihedral $D_2$ group, which includes a two-fold ($C_2$) rotational symmetry axis along the optical axis and two $C_2$ axes perpendicular to the optical axis. Based on the classification introduced by Menzel *et al.* [50, 58], the homogenized metamaterial slab thus exhibits elliptical counter-rotating eigenstates along the $\langle 110 \rangle$ direction. In contrast to what has been observed in "metallic gyroid photonic crystals" [59], none of these eigenstates are of the Bloch form in lossy metamaterials, but evanescent Floquet modes [60] with higher order surface modes involved in the scattering process [41]. Without additional approximations, symmetry implies the presence of a finite CD with symmetrical circular polarization conversion and a linear dichroism, as experimentally observed in [41]. The polarization conversion is further responsible for the CD observed in reflection (Figure S1) [4]. All experimental characterizations performed here are thus consistent with the symmetry properties of a $\langle 110 \rangle$ oriented gyroid slab.

A careful investigation of the spectral shapes reveal features that are not an immediate signature of the gyroid lattice: The minimum in the transmission spectra in Figure 4a at ca. 520 nm seems to be indicative of a strong resonance that is not predicted in any of the previous theories or simulation results [41,55]. However, due to the fabrication process of the the gyroid metamaterials, involving an electrodeposition process on rough FTO substrates (see Methods), surfaces on both sides of the gyroid slab show non-planar terminations.

The observed polarization-sensitive transmission dip may thus stem from the non-planar terminations of the fabricated gyroid samples that lead to chiral protrusions supporting associated localized plasmonic modes. Since the plasma resonance of silver is blue-shifted compared to gold, combined with the lower plasmonic losses in silver, a blue-shift of this resonance by approximately 100 nm and a substantial increase of its strength is expected in silver gyroids. In the absence of absorption at $\sim 400$ nm for any of the constituent materials, we deduce that the strong decay of the silver spectra below 500 nm has the same origin as the 520 nm minimum in gold.

A further interesting spectral feature in Figure 4b is the minimum at ca. 520 nm, which appears only in the LCP channel. Preliminary full wave simulations (not shown) indicate that this spectral feature arises from a special topography at one of the surfaces in the absence of bulk structural elements which provide filtering.

### 3.2 Tunability of the CD

The spectra in Figures 3 and 4 show that the CD is strongly influenced and tunable by the material composition. The fabrication process allows the replication of the copolymer template into different (plasmonic) materials, where the focus here is on silver and gold due to their favorable plasmonic properties at optical wavelengths [61]. Figure 3 shows that the CD of gold metamaterials is strongly reduced compared to silver metamaterials. The reduced CD can be associated with the higher intrinsic losses of the gold material properties [62]. The smaller losses of silver facilitates the resonant behavior of the nanostructure, resulting in a strong chiro-optical activity. The peak CD intensity is observed at $\sim 510$ nm for silver gyroids and, bathochromically shifted, at 550 nm for gold gyroids, consistent with the difference in the plasma frequency between the two materials [62].

A higher refractive index environment results in an increase in the transmission for both polarizations in gold at higher wavelengths above the resonance dip (Figure 4a). This arises from a shift of the surface plasmon polariton fields into the dielectric domain when the permittivity of the latter is increased, leading to decreased effective absorption. At the same time, the higher refractive index also increases the scattering cross-section of the localised resonance as expected from Mie theory, leading to decreased transmission at the resonance dip. The CD increases and red-shifts as seen in Figure 4c. The underlying mechanism seems to be a circular polarization sensitivity of the resonance at its long wavelength tail, indicating that it originates from chiral protrusions.

In silver gyroids, the optical response of the material depends strongly on the sample history. For the sample, in which the polymer was not removed, a similar result to gold was observed, with a marked increase in peak transmission in the RCP channel. Interestingly, the polymer filled gyroid leads to an increased LCP resonance dip at 520 nm, leaving the overall transmission in this channel approximately unchanged, resulting in a strongly enhanced CD (see Figure S2). For the samples that were voided in an Ar plasma and subsequently filled with dielectric oils, an opposite trend is observed - the transmission decreases with increasing refractive index in both channels. Since the optical difference between the polymer and oil-filled gyroids cannot be an effect of the dielectric constant alone, this must arise from differences in the sample preparation, i.e. the plasma removal of the polymer. As indicated above, there is evidence for the formation of a very thin carbonaceous layer at the silver surface during this process. From the spectra in Figure 4b, it is evident that this layer causes losses in the propagation of plasmon-polaritons across the gyroid network, thereby reducing the transmission in both channels. The overall dichroism of the oil-infiltrated samples is, however, increased compared to the air, which is indicative of a stronger transmission reduction in the LCP channel compared to the RCP signal.

### 3.3 Comparison to top-down methods

The strength of the CD is well above previously reported values for other materials created using self-assembly techniques [4] and equals CD values in metamaterials fabricated via top-down approaches: in a comparable wavelength range (450 - 550 nm), where only a few examples have been reported so far, the best CD results were achieved by Esposito *et al.* [30]. In their work, the transmitted light under CP illumination through intertwining triple-helical nanowires oriented along the optical axis has a CD contrast of ∼0.22 at 500 nm ($T_{\text{RCP}} = 0.51$, $T_{\text{LCP}} = 0.33$). A strong CD value is reported over a large transmission band, from 500 to 900 nm. Very recently, planar gammadion-shaped dielectric nanostructures [20] set the record CD value of 0.75 at 540 nm ($T_{\text{RCP}} = 0.85$, $T_{\text{LCP}} = 0.12$). However, in this work, the wavelength range of the CD is extremely narrow (from 530 to 550 nm). Our work demonstrates that 3D chiral silver nanostructures have a strong CD over a large wavelength range in the visible, comparable to the best results achieved via carefully designed top-down approaches. This strong CD between 450 and 550 nm, with a peak of 0.4 at 510 nm, has been found for fabricated gyroid samples that have a preferred ⟨110⟩ out-of-plane orientation. Comparable experiments on single gyroid metamaterials with other orientations would facilitate the understanding of the role of the metamaterial orientation on the optical properties. ⟨111⟩-oriented single gyroids have a higher degree of rotational symmetry ($C_3$) so that the associated eigenstates of the metamaterial seperate into the RCP and LCP channels, respectively [3, 50]. Therefore, a gyroid metamaterial oriented ⟨111⟩ out-of-plane would exhibit a CD in transmission but not in reflection, that is due to different absorption in the two channels, and zero polarization conversion. The absence of polarization conversion around an *n*-fold ($n>2$) symmetry [3] has been previously employed as a design principle to eliminate polarization conversion in top-down fabricated metamaterials [30] and in photonic crystals [3, 63]. The gyroid naturally exhibits 3-fold axes along its crystallographic ⟨111⟩ direction, so that a gyroid metamaterial with this out-of-plane orientation would be very interesting both from a fundamental and an application point-of-view. Gyroids oriented along their ⟨100⟩ direction are, on the other hand, expected to lead to a much stronger CD [55].

## 4 Conclusions

In conclusion, we have demonstrated the presence of a strong CD in metallic gyroid optical metamaterials. Through a block copolymer-based fabrication method, we fabricated silver and gold metamaterials exhibiting a single gyroid morphology with a cubic unit cell with ∼65 nm lattice constant. Optical micrographs and spectroscopic measurements with circularly polarized light reveal a strong CD between 470 and 570 nm. The material properties and the refractive index of the surrounding medium have a large influence on the CD. Compared to silver single gyroids, the CD in gold single gyroid metamaterials is significantly weaker and spectrally shifted. Both of these observations are expected from the larger losses and the frequency-shifted plasmonic response. Additionally, the CD is strongly affected by the host medium, where a higher refractive index results in a higher CD. The measured CD is much enhanced in comparison to structures previously fabricated by top-down lithography.

Our results further demonstrate an extreme sensitivity of the optical response to minute details in the 3D network morphology. The unexplained resonances in both the gold and silver spectra most likely arise from as yet unexplained surface resonances that have been shown to depend extremely sensitively on structural details at the gyroid surfaces. Particularly the spectral differences in the two CP channels in silver are puzzling and could stem from the presence of polarization-sensitive local resonances in chiral metamaterial protrusions. Further investigations are necessary to test this hypothesis. As expected for plasmonic nanostructures, the plasmonic excitations at the metal-dielectric interface within the metamaterial are highly sensitive to the particular interface composition. In the case of the silver gyroids studied here, this gives rise to a delicate interplay. While the passivation of the silver surface is essential to avoid its oxidation, the presence of carbonaceous layer at the silver surface is detrimental for the plasmonic performance. Creating a silver network within a polymer template (excluding air), on the other hand, provides excellent metamaterial properties without the need to additionally stabilize the silver.

Our results pave the way toward the bottom-up manufacture of 3D self-assembled silver optical metamaterials. In particular, the feasibility of electrochemical silver replication of polymeric 10-nm network morphologies demonstrates their utility for the creation of optical metamaterials. Applications of this process include tunable CD filters and a promising large-scale material for chiral sensing. Accessing other high symmetry orientations of the gyroid containing screw axes and associated helical elements is expected to generate even stronger chiro-optical properties, such as strong CD along the cubic $\langle 100 \rangle$ direction [55] and optical activity in the absence of ellipticity along the $C_3$ symmetric $\langle 111 \rangle$ direction [3].

# 5 Materials and Methods

**Sample Fabrication**

A 10% solution of polyisoprene-*b*-polystyrene-*b*-poly(ethylene oxide) (ISO) triblock terpolymer (80 kg/mol, $f_{\text{PI}} = 0.30$, $f_{\text{PS}} = 0.53$, $f_{\text{PEO}} = 0.17$) in anhydrous anisole (Sigma-Aldrich) was further spincoated for 60 s at 1200 rpm onto a fluorine-doped tin oxide coated glass substrate (FTO glass, Sigma-Aldrich). The FTO glass was immersed in a Piranha bath and then silanized by immersion in a 0.2% solution of octyltrichlorosilane (Sigma-Aldrich) in anhydrous anisole for 15 s. As-spun samples were annealed in a controlled solvent vapour atmosphere using an experimental set-up detailed in [40].

Annealed films were treated with UV light (Mineralight ® XX-15S, 254 nm) for 15 minutes and washed with ethanol absolute for 30 minutes to remove PI from the matrix and get the voided structure. The sample was backfilled by plasmonic material through electrodeposition. MetSil 500CNF (Metalor) silver solution and ECF 60 (Metalor) gold solution modified with arsenic trioxide were used for electrodeposition. The electrodeposition was carried out using a Metrohm AutoLab PGSTAT302N potentiostat. Ag/AgCl with KCl (Metrohm) and platinum electrode tip (Metrohm) were used as the reference and counter electrodes. The first step of the electrodeposition, the nucleation of the metallic structure on the conductive substrate, was performed using a start and lower vertex potential of -0.4 V and -1.15 V. The potential for the growing step was -0.756 V for gold and -0.656 V for silver.

To remove the polymer from the replica, plasma etching was used for 20 minutes (Plasma Etch Inc., PE-100-RIE, 50 W and 40CC/min Ar).

**Optical Characterization**

A Zeiss Axio Scope.A1 (Zeiss AG, Oberkochen, Germany) polarized light microscope and a xenon light source (Thorlabs SLS401; Thorlabs GmbH, Dachau, Germany) was used for all optical experiments. For spectroscopic measurements, the transmitted and reflected light was collected by an optical fibre (QP230-2-XSR, 230 µm core) with a measurement spot size of ≈20 µm. The spectra were recorded by a spectrometer (Ocean Optics Maya2000 Pro; Ocean Optics, Dunedin, FL, USA). Light microscopy images were acquired with a Point Grey GS3-U3-28S5C-C (FLIR Integrated Imaging Solutions Inc., Richmond, Canada) CCD camera. The linear polarizers (Thorlabs WP25M-UB, Thorlabs) and quarter waveplates (B. Halle, Germany) were both superachromatic, allowing an effective measurement range of 400 - 800 nm.

**Scanning electron microscopy (SEM)**

Samples were imaged using a TESCAN (TESCAN, a.s., Brno, Czech Republic) MIRA3 field emission scanning electron microscope.

**Supporting Information**

**Acknowledgements**

This research was supported by the Swiss National Science Foundation through grant number 163220and 188647, the Ambizione program (168223 to BDW), and the National Centre of Competence in Research "Bio-Inspired Materials".

**Conflict of Interest**

The authors declare no conflict of interest.

## Keywords

circular dichroism, chiral media, optical metamaterials, 3D nanostructure

Supplementary information for

# Strong Circular Dichroism in Single Gyroid Optical Metamaterials

Cédric Kilchoer, Narjes Abdollahi, James A. Dolan, Doha Abdelrahman, Matthias Saba, Ulrich Wiesner, Ullrich Steiner, Ilja Gunkel and Bodo D. Wilts

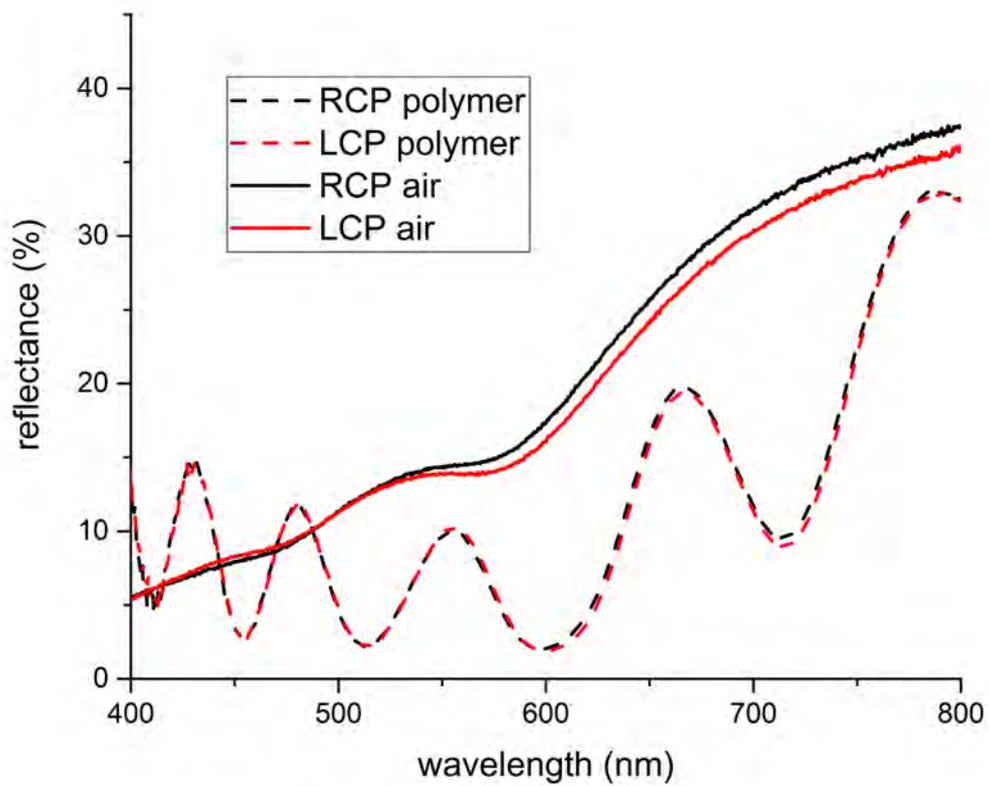

*Figure S1 : Circular dichroism in reflection under RCP and LCP illumination with polymer and air as surrounding medium.*

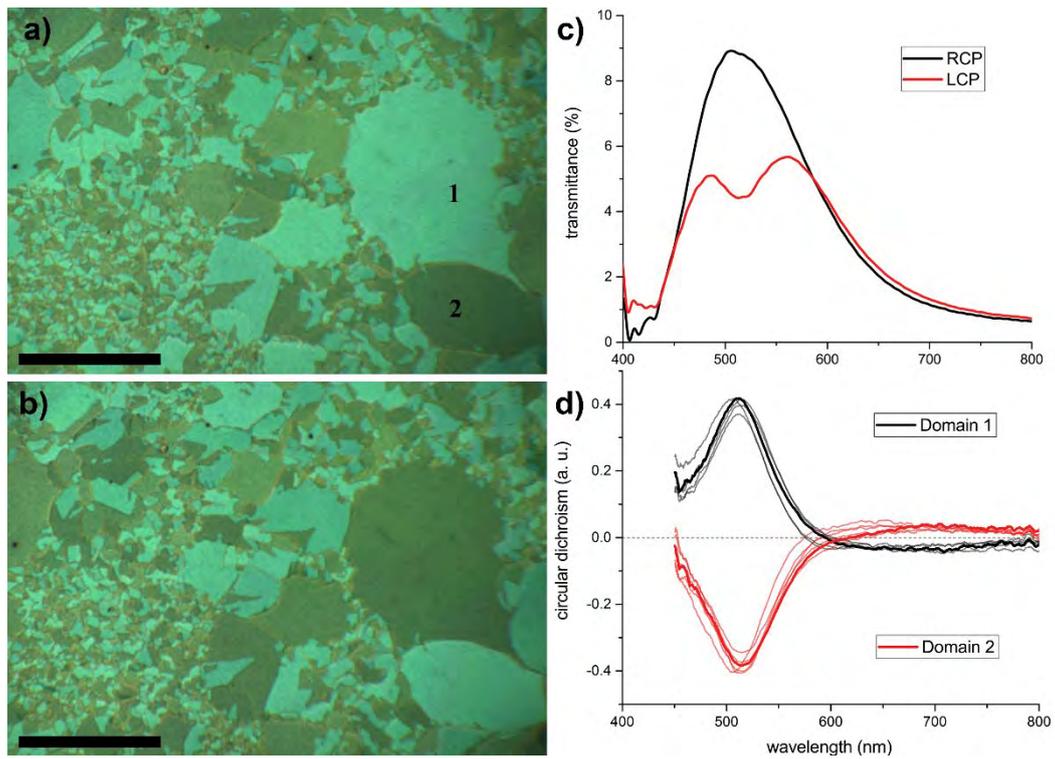

*Figure S2: Circular dichroism in silver gyroid optical metamaterial with polymer as surrounding medium. Samples are illuminated by right-handed (a) and left-handed (b) circularly polarized light. Scale bar is 500 µm. (c) Transmittance spectra of an individual domain collected after the RCP and LCP filtering. (d) CD measured on 10 different domains (d). The circular dichroism of domains 1 and 2, marked in (a), is highlighted.*

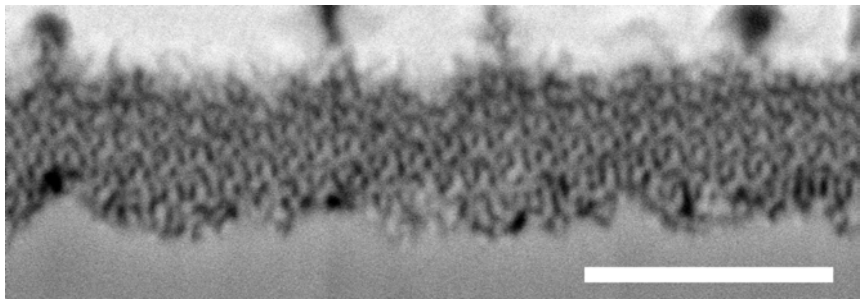

*Figure S3: Cross-sectional view of a silver gyroid metamaterial using FIB-SEMl. The nanostructure is periodic and well-ordered throughout the film thickness. Scale bar: 500 nm.*